\documentclass[aps,prl,preprint]{revtex4}
\usepackage{graphicx,color}
\usepackage{dcolumn}
\usepackage{bm}
\usepackage{amsmath,amsthm,amssymb}

\begin{document}

\def\average#1{\left\langle {#1} \right\rangle}
\def\averagefr#1{\left\langle {#1} \right\rangle_0}
\def\avbar#1{\overline{#1}}
\def\impaverage#1{\left\langle {#1} \right\rangle_{\rm i}}
\def\bra#1{\lt\langle{#1} \rt|}
\def\ket#1{\lt|{#1} \rt\rangle}
\def\braket#1#2{\lt.\lt\langle{#1} \rt|{#2}\rt\rangle}
\def\ddelo#1{\frac{d^2}{d #1^2}}
\def\ddel#1#2{\frac{d^2 #1}{d #2 ^2}}
\def\ddelpo#1{\frac{\partial^2}{\partial #1^2}}
\def\delo#1{\frac{d}{d #1}}
\def\delpo#1{\frac{\partial}{\partial #1}}
\def\deldo#1{\frac{\delta}{\delta #1}}
\def\del#1#2{\frac{d #1}{d #2}}
\def\delp#1#2{\frac{\partial #1}{\partial #2}}
\def\deld#1#2{\frac{\delta #1}{\delta #2}}
\def\vvec#1{\stackrel{{\leftrightarrow}}{#1}} 
\def\vectw#1#2{\left(\begin{array}{c} #1 \\ #2 \end{array}\right)}
\def\vecth#1#2#3{\left(\begin{array}{c} #1 \\ #2 \\ #3 \end{array}\right)}
\def\mattw#1#2#3#4{\left(\begin{array}{cc} #1 & #2 \\ #3 & #4 \end{array}\right)}
\def\Eqref#1{Eq. (\ref{#1})}
\def\Eqsref#1#2{Eqs. (\ref{#1})(\ref{#2})}
\def\Eqrefj#1{(\ref{#1})式}
\def\Eqsrefj#1#2{式 (\ref{#1})(\ref{#2})}
\def\dispeq#1{$\displaystyle{#1}$}
\def\listitem#1{\begin{itemize}\item #1 \end{itemize}}
\newcommand{\lt}{\left}
\newcommand{\rt}{\right}
\newcommand{\nablarl}{\stackrel{\leftrightarrow}{\nabla}}
\newcommand{\nablar}{\stackrel{\rightarrow}{\nabla}}
\newcommand{\nablal}{\stackrel{\leftarrow}{\nabla}}
\newcommand{\nnr}{\nonumber\\}
\newcommand{\adag}{{a^{\dagger}}}
\newcommand{\alphaz}{\alpha_0}
\newcommand{\alphasf}{\alpha_\spinflip}
\newcommand{\Area}{A}
\newcommand{\aT}{\overline{\rm T}}
\newcommand{\av}{{\bm a}}
\newcommand{\Av}{{\bm A}}
\newcommand{\Avem}{{\bm A}_{\rm em}}
\newcommand{\Aem}{{A}_{\rm em}}
\newcommand{\Ams}{{\rm A/m}^2}
\newcommand{\Aph}{A^{\phi}}
\newcommand{\Ath}{A^{\theta}}
\newcommand{\Aphv}{\Av^{\phi}}
\newcommand{\Athv}{\Av^{\theta}}
\newcommand{\Az}{{A^{z}}}
\newcommand{\bv}{{\bm b}}
\newcommand{\Bv}{{\bm B}}
\newcommand{\Bz}{{B_z}}
\newcommand{\Bc}{B_{\rm c}}
\newcommand{\Bvs}{{\bm B}_{S}}
\newcommand{\Bveff}{{\bm B}_{\rm eff}}
\newcommand{\Bve}{{\bm B}_{\rm e}}
\newcommand{\Bwb}{B_{\rm w}}
\newcommand{\betasf}{{\beta_\spinflip}}
\newcommand{\betana}{{\beta_{\rm na}}}
\newcommand{\betaw}{{\beta_{\rm w}}}
\newcommand{\cbar}{\avbar{c}}
\newcommand{\cdag}{{c^{\dagger}}}
\newcommand{\chiz}{\chi^{(0)}}
\newcommand{\chio}{\chi^{(1)}}
\newcommand{\chitilo}{\tilde{\chi}^{(1)}}
\newcommand{\chitilz}{\tilde{\chi}^{(0)}}
\newcommand{\chiuni}{\chi_{0}}
\newcommand{\cH}{c_{H}}
\newcommand{\cHdag}{c_{H}^{\dagger}}
\newcommand{\ckv}{c_{\kv}}
\newcommand{\ckvs}{c_{\kv\sigma}}
\newcommand{\ccv}{{\bm c}}
\newcommand{\Cbeta}{C_\beta}
\newcommand{\Cr}{C_\rightarrow}
\newcommand{\Cl}{C_\leftarrow}
\newcommand{\Ci}{C_{\rm i}}
\newcommand{\Ct}{C_t}
\newcommand{\Cv}{{\bm C}}
\newcommand{\cv}{{\bm c}}
\newcommand{\ctil}{\tilde{c}}
\newcommand{\dbar}{\avbar{d}}
\newcommand{\deltaS}{\delta S}
\newcommand{\Deltasd}{\Delta_{sd}}
\newcommand{\dels}{{s_0}}
\newcommand{\ddagg}{d^{\dagger}}
\newcommand{\dtil}{\tilde{d}}
\newcommand{\dw}{{\rm w}}
\newcommand{\dx}{{d^3 x}}
\newcommand{\Deltatil}{\tilde{\Delta}}
\newcommand{\Dcal}{{\cal D}}
\newcommand{\DOS}{N}
\newcommand{\dos}{\nu}
\newcommand{\doss}{\nu_0}
\newcommand{\dosz}{\nu_0}
\newcommand{\DOSV}{{N(0)}}
\newcommand{\DOSom}{{\DOS_\omega}}
\newcommand{\ef}{{\epsilon_F}}
\newcommand{\efd}{{\epsilon_F^{(d)}}}
\newcommand{\eF}{{\epsilon_F}}
\newcommand{\eft}{{\epsilon_F \tau}}
\newcommand{\eftauinv}{\frac{\hbar}{\epsilon_F \tau}}
\newcommand{\edk}{\epsilon_{\kv}^{(d)}}
\newcommand{\ekv}{\epsilon_{\kv}}
\newcommand{\ekvs}{\epsilon_{\kv\sigma}}
\newcommand{\elld}{{\ell_{\rm D}}}
\newcommand{\Ev}{{\bm E}}
\newcommand{\ev}{{\bm e}}
\newcommand{\evth}{{\bm e}_{\theta}}
\newcommand{\evph}{{\bm e}_{\phi}}
\newcommand{\evs}{{\bm n}}
\newcommand{\evsz}{{\evs}_0}
\newcommand{\evsph}{(\evph\times\evz)}
\newcommand{\evx}{{\bm e}_{x}}
\newcommand{\evy}{{\bm e}_{y}}
\newcommand{\evz}{{\bm e}_{z}}
\newcommand{\eoh}{\frac{e}{\hbar}}
\newcommand{\fl}{{\eta}}
\newcommand{\fltil}{\tilde{\eta}}
\newcommand{\flitil}{\tilde{\fl_{\rm I}}}
\newcommand{\flrtil}{\tilde{\fl_{\rm R}}}
\newcommand{\fli}{{\fl_{\rm I}}}
\newcommand{\flr}{{\fl_{\rm R}}}
\newcommand{\fkv}{f_{\kv}}
\newcommand{\fkvs}{f_{\kv\sigma}}
\newcommand{\fo}{{f(\omega)}}
\newcommand{\fpo}{{f'(\omega)}}
\newcommand{\fomega}{{\omega}}
\newcommand{\fbeta}{f^{\beta}}
\newcommand{\fpin}{{f_{\rm pin}}}
\newcommand{\Fpin}{{F_{\rm pin}}}
\newcommand{\fe}{{f_{\rm e}}}
\newcommand{\fna}{{f_{\rm ref}}}
\newcommand{\Fe}{F}
\newcommand{\Fev}{\Fv}
\newcommand{\Fbeta}{F^{\beta}}
\newcommand{\Fbetav}{\Fv^{\beta}}
\newcommand{\Fbetafactor}{\mu}
\newcommand{\Fhall}{F^{\rm Hall}}
\newcommand{\Fhallv}{\Fv^{\rm Hall}}
\newcommand{\Fren}{F^{\rm ren}}
\newcommand{\Frenv}{\Fv^{\rm ren}}
\newcommand{\Fvna}{\Fv^{\rm ref}}
\newcommand{\Fnav}{\Fv^{\rm ref}}
\newcommand{\Fna}{F^{\rm ref}}
\newcommand{\Fad}{\Fhall}
\newcommand{\Fzad}{F^{\rm (0)ad}}
\newcommand{\Fv}{{\bm F}}
\newcommand{\Fo}{F^{(1)}}
\newcommand{\Fw}{F_{\rm w}}
\newcommand{\Fz}{F^{(0)}}
\newcommand{\Fzv}{\Fv^{(0)}}
\newcommand{\Fdelta}{\delta F}
\newcommand{\Fdeltav}{\delta \Fv}
\newcommand{\Fdel}{\delta \Fo}
\newcommand{\Fdelv}{\delta \Fv^{(1)}}
\newcommand{\ftil}{\tilde{f}}
\newcommand{\Gv}{{\bm G}}
\newcommand{\gv}{{\bm g}}
\newcommand{\gr}{g^{\rm r}}
\newcommand{\gto}{g^{\rm t}}
\newcommand{\ga}{g^{\rm a}}
\newcommand{\Gr}{G^{\rm r}}
\newcommand{\Ga}{G^{\rm a}}
\newcommand{\Gto}{G^{\rm t}}
\newcommand{\Gat}{G^{\overline{\rm t}}}
\newcommand{\Gless}{G^{<}}
\newcommand{\Ggrt}{G^{>}}
\newcommand{\gless}{g^{<}}
\newcommand{\ggrt}{g^{>}}
\newcommand{\Gtil}{\tilde{G}}
\newcommand{\Gcal}{{\cal G}}
\newcommand{\gap}{\Delta_{\rm sw}}
\newcommand{\gammap}{\gamma_{+}}
\newcommand{\gammam}{\gamma_{-}}
\newcommand{\gammaz}{\gamma_{0}}
\newcommand{\grst}{|0\rangle}
\newcommand{\grsthc}{\langle0|}
\newcommand{\grft}{|\ \rangle}
\newcommand{\gyro}{\gamma}
\newcommand{\gyroz}{\gyro_{0}}
\newcommand{\hf}{\frac{1}{2}}
\newcommand{\HA}{{H_{A}}}
\newcommand{\HB}{H_{B}}
\newcommand{\Hv}{\bm{H}}
\newcommand{\He}{H_{\rm e}}
\newcommand{\Heff}{H_{\rm eff}}
\newcommand{\Hem}{H_{\rm em}}
\newcommand{\Hex}{H_{\rm ex}}
\newcommand{\Himp}{H_{\rm imp}}
\newcommand{\Hint}{H_{\rm int}}
\newcommand{\HR}{{H_{\rm R}}}
\newcommand{\Hs}{{H_{\rm S}}}
\newcommand{\Hsf}{{H_\spinflip}}
\newcommand{\Hso}{{H_{\rm so}}}
\newcommand{\Hsd}{H_{sd}}
\newcommand{\Hst}{{H_{\rm ST}}}
\newcommand{\Hw}{H_{\dw}}
\newcommand{\Hz}{H_{0}}
\newcommand{\hbarinv}{\frac{1}{\hbar}}
\renewcommand{\Im}{{\rm Im}}
\newcommand{\ime}{\gamma}
\newcommand{\intinf}{\int_{-\infty}^{\infty}}
\newcommand{\intek}{\int_{-\ef}^{\infty}\! d\epsilon}
\newcommand{\intom}{\int\! \frac{d\omega}{2\pi}}
\newcommand{\intx}{\int\! {d^3x}}
\newcommand{\intk}{\int\! \frac{d^3k}{(2\pi)^3}}
\newcommand{\intr}{\int\! {d^3r}}
\newcommand{\intt}{\int_{-\infty}^{\infty}\! {dt}}
\newcommand{\ioh}{\frac{i}{\hbar}}
\newcommand{\iv}{\bm{i}}
\newcommand{\Ibar}{\overline{I}}
\newcommand{\Iv}{\bm{I}}
\newcommand{\Jex}{{J_{\rm ex}}}
\newcommand{\Jsd}{J_{sd}}
\newcommand{\Js}{{J_{\rm s}}}
\newcommand{\Jv}{\bm{J}}
\newcommand{\ja}{j_{\rm a}}
\newcommand{\js}{j_{\rm s}}
\newcommand{\jsc}{{j_{\rm s}^{\rm c}}}
\newcommand{\jsv}{\bm{j}_{\rm s}}
\newcommand{\Jsv}{\bm{J}_{\rm S}}
\newcommand{\JSv}{\bm{J}_{\rm S}}
\newcommand{\JS}{J_{\rm S}}
\newcommand{\JStotv}{\bm{J}_{S,{\rm tot}}}
\newcommand{\jc}{j_{\rm c}}
\newcommand{\jci}{{{j}_{\rm c}^{\rm i}}}
\newcommand{\jce}{{{j}_{\rm c}^{\rm e}}}
\newcommand{\jatil}{{\tilde{j}_{\rm a}}}
\newcommand{\jctil}{{\tilde{j}_{\rm c}}}
\newcommand{\jcitil}{{\tilde{j}_{\rm c}^{\rm i}}}
\newcommand{\jcetil}{{\tilde{j}_{\rm c}^{\rm e}}}
\newcommand{\jstil}{{\tilde{j}_{\rm s}}}
\newcommand{\jtil}{{\tilde{j}}}
\newcommand{\jv}{\bm{j}}
\newcommand{\kB}{{k_B}}
\newcommand{\kb}{{k_B}}
\newcommand{\kv}{{\bm k}}
\newcommand{\kvxv}{\kv\cdot\xv}
\newcommand{\kvo}{{\kv_1}}
\newcommand{\kvp}{{\kv}'}
\newcommand{\kpq}{{k+\frac{q}{2}}}
\newcommand{\kmq}{{k-\frac{q}{2}}}
\newcommand{\kvpq}{{\kv}+\frac{\qv}{2}}
\newcommand{\kvmq}{{\kv}-\frac{\qv}{2}}
\newcommand{\kvopq}{{\kvo}+\frac{\qv}{2}}
\newcommand{\kvomq}{{\kvo}-\frac{\qv}{2}}
\newcommand{\kvppq}{{\kvp}+\frac{\qv}{2}}
\newcommand{\kvpmq}{{\kvp}-\frac{\qv}{2}}
\newcommand{\kf}{{k_F}}
\newcommand{\kF}{{k_F}}
\newcommand{\kfpm}{{k_{F\pm}}}
\newcommand{\kfmp}{{k_{F\mp}}}
\newcommand{\kfu}{k_{F+}}
\newcommand{\kfd}{k_{F-}}
\newcommand{\kfs}{k_{F\sigma}}
\newcommand{\kfdel}{k_{F}^{(d)}}
\newcommand{\kom}{{k_\omega}}
\newcommand{\Kp}{{K_\perp}}
\newcommand{\kpv}{{\kv_\perp}}
\newcommand{\ktil}{\tilde{k}}
\newcommand{\lambdaD}{\lambda_{\rm D}}
\newcommand{\lams}{{\lambda_{\rm s}}}
\newcommand{\lamv}{{\lambda_{\rm v}}}
\newcommand{\lamso}{{\lambda_{\rm so}}}
\newcommand{\lamz}{{\lambda_{0}}}
\newcommand{\lambdaperp}{{\lambda_{\perp}}}
\newcommand{\Lcal}{{\cal L}}
\newcommand{\Le}{{L_{\rm e}}}
\newcommand{\Lez}{{L_{\rm e}^0}}
\newcommand{\Leff}{L_{\rm eff}}
\newcommand{\Lb}{L_{\rm B}}
\newcommand{\Ldw}{L_{\dw}}
\newcommand{\Lsd}{{L_{sd}}}
\newcommand{\Ls}{{L_{S}}}
\newcommand{\Lsw}{L_{\rm sw}}
\newcommand{\Lswdw}{L_{\rm sw-dw}}
\newcommand{\Linv}{{\frac{1}{L}}}
\newcommand{\lstil}{\tilde{l_\sigma}}
\newcommand{\Lv}{\bm{L}}
\newcommand{\md}{m^{(d)}}
\newcommand{\mv}{{\bm m}}
\newcommand{\Mv}{{\bm M}}
\newcommand{\Mphi}{{M_{\phi}}}
\newcommand{\Mw}{{M_{\dw}}}
\newcommand{\Ms}{M_{\rm s}}
\newcommand{\MR}{{\rm MR}}
\newcommand{\Mz}{M}
\newcommand{\mus}{{g\mu_{B}}}
\newcommand{\mub}{\mu_B}
\newcommand{\muB}{\mu_B}
\newcommand{\muz}{\mu_0}
\newcommand{\nel}{n_{\rm e}}
\newcommand{\Ne}{N_{\rm e}}
\newcommand{\nv}{{\bm n}}
\newcommand{\Nimp}{N_{\rm imp}}
\newcommand{\nimp}{n_{\rm imp}}
\newcommand{\Ninv}{\frac{1}{N}}
\newcommand{\Nw}{N_{\dw}}
\newcommand{\nvortex}{n_{\rm v}}
\newcommand{\nvz}{{\nv}_0}
\newcommand{\nz}{{n}_0}
\newcommand{\om}{{\omega}}
\newcommand{\omegap}{\omega'}
\newcommand{\Omegatil}{\tilde{\Omega}}
\newcommand{\Omegap}{\Omega'}
\newcommand{\Omegapin}{\Omega_{\rm pin}}
\newcommand{\ompOm}{{\omega+\frac{\Omega}{2}}}
\newcommand{\ommOm}{{\omega-\frac{\Omega}{2}}}
\newcommand{\Omz}{\Omega_0}
\newcommand{\ompOmz}{{\omega+\frac{\Omz}{2}}}
\newcommand{\ommOmz}{{\omega-\frac{\Omz}{2}}}
\newcommand{\Omhf}{\frac{\Omega}{2}}
\newcommand{\phiso}{\phi_{\rm so}}
\newcommand{\phiz}{{\phi_0}}
\newcommand{\Phiv}{\bm{\Phi}}
\newcommand{\PhiB}{{\Phi_{\rm B}}}
\newcommand{\Ptil}{{\tilde{P}}}
\newcommand{\pv}{{\bm p}}
\newcommand{\PDOS}{P_{\DOS}}
\newcommand{\qv}{{\bm q}}
\newcommand{\qvxv}{\qv\cdot\xv}
\newcommand{\qtil}{{\tilde{q}}}
\newcommand{\ra}{\rightarrow}
\renewcommand{\Re}{{\rm Re}}
\newcommand{\rhoB}{{\rho_{\rm B}}}
\newcommand{\rhob}{{\rho_{\rm B}}}
\newcommand{\rhow}{{\rho_{\dw}}}
\newcommand{\rhos}{{\rho_{\rm s}}}
\newcommand{\rhoS}{{\rho_{\rm s}}}
\newcommand{\rhoxy}{{\rho_{xy}}}
\newcommand{\RS}{{R_{\rm S}}}
\newcommand{\Rw}{{R_{\dw}}}
\newcommand{\rv}{{\bm r}}
\newcommand{\Rv}{{\bm R}}
\newcommand{\sd}{$s$-$d$}
\newcommand{\sigmav}{{\bm \sigma}}
\newcommand{\sigmab}{\sigma_{\rm B}}
\newcommand{\sigmaB}{\sigma_{\rm B}}
\newcommand{\sigmaz}{\sigma_0}
\newcommand{\se}{{s}}
\newcommand{\sev}{{\bm \se}}
\newcommand{\sevsf}{{\bm \se}_\spinflip}
\newcommand{\SE}{\Sigma}
\newcommand{\SEr}{\Sigma^{\rm r}}
\newcommand{\SEa}{\Sigma^{\rm a}}
\newcommand{\SEless}{\Sigma^{<}}
\newcommand{\sgn}{{\rm sgn}}
\newcommand{\sz}{{s}_0}
\newcommand{\sv}{{{\bm s}}}
\newcommand{\seth}{{\se}_\theta}
\newcommand{\seph}{{\se}_\phi}
\newcommand{\sez}{{\se}_z}
\newcommand{\so}{{\rm so}}
\newcommand{\spol}{\Deltasd}
\newcommand{\spinflip}{{\rm sr}}
\newcommand{\svtil}{\tilde{\bm s}}
\newcommand{\stil}{\tilde{\se}}
\newcommand{\stilz}{\stil_{z}}
\newcommand{\stilpm}{\stil^{\pm}}
\newcommand{\stilpmz}{\stil^{\pm(0)}}
\newcommand{\stilpma}{\stil^{\pm(1{\rm a})}}
\newcommand{\stilpmb}{\stil^{\pm(1{\rm b})}}
\newcommand{\stilpmo}{\stil^{\pm(1)}}
\newcommand{\stilpmv}{\stil^{\pm(V)}}
\newcommand{\stilpara}{\stil_{\parallel}}
\newcommand{\stilperp}{\stil_{\perp}}
\newcommand{\Simpv}{{{\bm S}_{\rm imp}}}
\newcommand{\Simp}{{S_{\rm imp}}}
\newcommand{\Ssd}{\overline{S}}
\newcommand{\Stot}{\Ssd}
\newcommand{\Stotv}{\bm{S}_{\rm tot}}
\newcommand{\Sh}{{\hat {S}}}
\newcommand{\Svh}{{\hat {\Sv}}}
\newcommand{\Sv}{{{\bm S}}}
\newcommand{\Svz}{{{\bm S}_0}}
\newcommand{\sumx}{{\int\! \frac{d^3x}{a^3}}}
\newcommand{\sumk}{{\sum_{k}}}
\newcommand{\sumkv}{{\sum_{\kv}}}
\newcommand{\sumqv}{{\sum_{\qv}}}
\newcommand{\sumom}{\int\!\frac{d\omega}{2\pi}}
\newcommand{\sumOm}{\int\!\frac{d\Omega}{2\pi}}
\newcommand{\sumomOm}{\int\!\frac{d\omega}{2\pi}\int\!\frac{d\Omega}{2\pi}}
\newcommand{\td}{t^{(d)}}
\newcommand{\thickness}{d}
\newcommand{\thetaz}{{\theta_0}}
\newcommand{\tr}{{\rm tr}}
\newcommand{\To}{{\rm T}}
\newcommand{\Ta}{\overline{\rm T}}
\newcommand{\Tc}{{\rm T}_{C}}
\newcommand{\Tct}{{\rm T}_{\Ct}}
\newcommand{\tcmp}{\tau}
\newcommand{\tcmpi}{\tau_{\rm I}}
\newcommand{\tcmpinf}{\tau_{\infty}}
\newcommand{\tcmpz}{\tau_{0}}
\newcommand{\tcmpzp}{\tau_{0}'}
\newcommand{\Torqv}{{\bm \tau}}
\newcommand{\torque}{{\tau}}
\newcommand{\torquej}{{\tau}_j}
\newcommand{\torquev}{{\bm \torque}}
\newcommand{\Torquev}{\torquev} 
\newcommand{\torqueve}{\torquev}
\newcommand{\torquee}{\torque}
\newcommand{\torquetil}{\tilde{\torque}}
\newcommand{\torquew}{{\torque_{\dw}}}
\newcommand{\tautil}{{\tilde{\tau}}}
\newcommand{\taup}{\tau_{+}}
\newcommand{\taum}{\tau_{-}}
\newcommand{\tauw}{\tau_{\dw}}
\newcommand{\tausf}{\tau_\spinflip}
\newcommand{\tauso}{\tau_{\rm(so)}}
\newcommand{\thetast}{\theta_{\rm st}}
\newcommand{\ttil}{{\tilde{t}}}
\newcommand{\tinf}{t_\infty}
\newcommand{\tz}{t_0}
\newcommand{\Ubar}{\overline{U}}
\newcommand{\Ueff}{U_{\rm eff}}
\newcommand{\Uz}{U_0}
\newcommand{\Uv}{U_V}
\newcommand{\vc}{{v_{\rm c}}}
\newcommand{\ve}{{v_{\rm e}}}
\newcommand{\vev}{{\vv_{\rm e}}}
\newcommand{\vv}{\bm{v}}
\newcommand{\vs}{{v_{\rm s}}}
\newcommand{\vsv}{{\vv_{\rm s}}}
\newcommand{\vw}{v_{\rm w}}
\newcommand{\vf}{{v_F}}
\newcommand{\vimp}{v_{\rm imp}}
\newcommand{\vi}{{v_{\rm i}}}
\newcommand{\Vi}{{V_{\rm i}}}
\newcommand{\Vso}{v_{\rm so}}
\newcommand{\vtil}{{{v_0}}}
\newcommand{\Vpin}{{V}_{\rm pin}}
\newcommand{\Vinv}{\frac{1}{V}}
\newcommand{\vwadiabatic}{v_{\rm ST}}
\newcommand{\vz}{{v_0}}
\newcommand{\Vz}{{V_0}}
\newcommand{\Vcal}{{\cal V}}
\newcommand{\Vztil}{{\tilde{V_0}}}
\newcommand{\varphiv}{\bm{\varphi}}
\newcommand{\varphiz}{\varphi_0}
\newcommand{\Ws}{{W_{\rm S}}}
\newcommand{\Xtil}{{\tilde{X}}}
\newcommand{\xv}{{\bm x}}
\newcommand{\xpv}{{\xv_\perp}}
\newcommand{\Xv}{{\bm X}}
\newcommand{\xvp}{{\bm x}_{\perp}}
\newcommand{\xw}{{z}}
\newcommand{\Xz}{{X_0}}
\newcommand{\Zs}{Z_{\rm S}}
\newcommand{\Zz}{Z_{0}}
\newcommand{\ztil}{u}
\newcommand{\zh}{\hat{z}}


\title{
Spin-transfer torque in disordered weak ferromagnets
}

\author{Yoshisuke Ban\footnote{
Present address:  Department of Electrical Engineering and Information Systems,
The University of Tokyo,
7-3-1 Hongo, Bunkyo-ku, Tokyo 113-8656, Japan.
}, 
Gen Tatara}
\affiliation{%
Department of Physics, Tokyo Metropolitan University,
Hachioji, Tokyo 192-0397, Japan\\
}

\date{\today}

\begin{abstract}
We study theoretically the spin transfer effect on a domain wall in disordered weak ferromagnets.
We have identified the adiabatic condition for the disordered case as
$\lambda \gg  \lambda_{\rm D}\equiv \sqrt{{\hbar D}/{\spol}}$, where $D$ and $\spol$ are the diffusion constant and the spin splitting energy due to the $s$-$d$ type exchange interaction, respectively, and found out that perfect spin-transfer effect occurs even in weak ferromagnets as long as this condition is satisfied.
The effective $\beta$ term arising from the force turns out to govern the wall dynamics, and therefore, 
the wall motion can be as efficient as in strong ferromagnets even if $\spol$ is small.
\end{abstract}


\maketitle

The spin-transfer effect has been intensively studied recently as a novel efficient magnetization switching mechanism without magnetic field.
The effect arises from the transfer of the spin angular momentum between the conduction electron and the localized spins (magnetization), as first pointed out by Berger in the case of domain wall \cite{Berger86}.
In conventional 3$d$ ferromagnets, domain walls are thick (with thickness $\lambda \sim 100$nm) and the coupling between the electron spin and the localized spin ($s$-$d$ type interaction) 
is strong. 
This is the adiabatic regime where the electron spin can follow the localized spin texture when traversing the domain wall. 
In the ballistic case, the adiabatic condition is given by \cite{Waintal04}
\begin{align}
\lambda \gg \frac{\vf}{\spol},
\end{align}
where $\lambda$ is the wall thickness, $\vf$ is the Fermi velocity and $\spol$ is the spin splitting of the conduction electron.
In this limit, the motion of the wall under current is dominated by the spin-transfer effect and the efficiency is governed by the angular momentum conservation law \cite{Berger86}.
The microscopic justification for the series of pionnering works by Berger \cite{Berger78,Berger86} was done in Refs. \cite{TK04,TKS_PR08}.

The effect of spin-relaxation, which acts as an effective non-adiabaticity, was studied by Zhang and Li and Thiaville et al. \cite{Zhang04,Thiaville05} and was shown to significantly affect the current-driven wall motion close to the adiabatic limit. 
The effect of non-adiabaticity (electron reflection by the wall) studied by Berger \cite{Berger78} and one of the present authors \cite{TK04} was then identified as another crucial factor even in the case close to the adiabatic limit.

While diffusive electron transport has been considered in the context of domain wall resistance \cite{TF97,Falloon06} 
and tunnel junctions \cite{Theodoropoulou07},
only ballistic case has been considered in discussing the spin-transfer torque.
The reason would be that normal (spin-conserving) impurity scattering has been believed not to affect the spin transfer processes.
In contrast, spin relaxation due to the spin-dependent impurity scattering and the spin-orbit interaction was shown to give rise to the effective force (called $\beta$ term) and thus modifies the domain wall motion greatly \cite{Tserkovnyak06,KTS06,TE08}.

\begin{figure}[tbh]
\begin{center}
\includegraphics[scale=0.7]{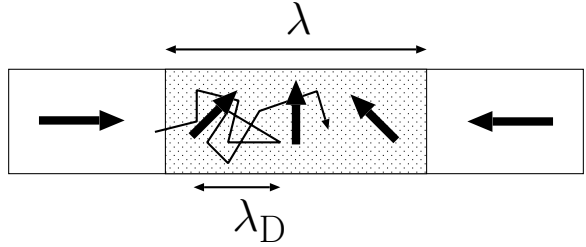}
\caption{ 
The diffusive adiabatic limit we consider. 
$\lambdaD=\sqrt{D\hbar/\spol}$ is the lenth scale the diffusive electron can reach within the precession period of its spin, $\hbar/\spol$. 
If the wall thickness is larger than $\lambdaD$, the electron spin can follow the localized spin profile and adiabatic spin transfer occurs.
\label{FIGdiffusivemotion}
}
\end{center}
\end{figure}

The aim of the present paper is to investigate the effect of normal impurities on the spin transfer effects, including the diffusion ladder.
We demonstrate that the diffusive electron motion results in the modification of the adiabatic condition to be
\begin{align}
\lambda \gg \sqrt{ \frac{\hbar D}{\spol} } \equiv\lambdaD ,
\label{adiabaticdisorder}
\end{align}
where $D=\frac{\hbar^2\kf^2}{3m^2}\tau$ is the diffusion constant with $\kf$, $m$ and $\tau$ being the Fermi energy, electron mass and elastic lifetime, respectively.
Here, $\lambdaD$ is the distance that a electron can reach diffusively within a time of spin precession caused by the $s$-$d$ interaction, $\spol$. 
Therefore if disordered adiabatic condition is satisfied, the electron spin can follow the localized spin profile while going through the wall if $\lambda \gg \lambdaD$ (Fig. \ref{FIGdiffusivemotion}). 
As a result, even in a weak ferromagnets with $\spol \ll \ef$ ($\ef$ is the Fermi energy), 
perfect spin transfer effect is realized if the system is disordered.
The adiabatic condition 
(\ref{adiabaticdisorder}) was already pointed out in Ref. \cite{TKS_PR08}, but the effect of diffusion ladder was not discussed there since only the strong spin splitting case was considered there.

The Lagrangian we consider is given by \cite{TKS_PR08}
\begin{eqnarray}
\Lez\equiv \hbarinv \intx \left[ i\hbar \cdag \dot{c} 
-  \left( \frac{\hbar^2}{2m}|\nabla c|^2 -\eF c^\dagger c \right)
+ {\spol}
  \evs \cdot (c^\dagger \sigmav c) \right]+\Himp,
  \label{Le0}
\end{eqnarray}
where $\nv(\xv)$ represents the direction of the lozalized spin.
In this paper, we consider the case of a planar wall, given by  
$n_z(\xv)=\tanh \frac{\xw}{\lambda}$, 
$n_x(\xv) \pm i n_y(\xv)= e^{\pm i \phi} \frac{1}{\cosh \frac{\xw}{\lambda} }$, 
where $\xw$ is the direction perpendicular to the wall plane and $\phi$ is a constant representing the angle out of the easy plane.
Scattering by normal impurities is described by $\Himp$. 
Treating the impurity potential as an on-site type, it is given by
\begin{eqnarray}
\Himp &=& \sum_{i=1}^{\Nimp}\sum_{\kv\kv'} 
\frac{\vimp}{N} e^{i(\kv-\kv')\cdot\Rv_i} \cdag_{\kv'}c_{\kv},
\end{eqnarray}
where $\vimp$ represents the strength of the impurity potential, $\Rv_i$ represents the position of random impurities, $\Nimp$ is the number of impurities, and 
$N\equiv V/a^3$ is number of sites.
To estimate physical quantities, we take the random average over impurity positions.
The self-energy type processes due to the impurity scattering results in the electron Green's function with lifetime $\tau$, e.g.,
$\gr_{\kv}(\omega)=\frac{1}{\omega-\ekv+\frac{i}{2\tau}}$, 
where the inverse lifetime is given as
\begin{equation}
\frac{1}{\tau} =
\frac{2\pi}{\hbar}{\nimp} \vimp^2 \dos.
\end{equation}
Here $\dos$ is the density of states per site at the Fermi level and
$\nimp\equiv \frac{\Nimp}{N}$ is the impurity concentration.
In this paper, we consider a weak ferromagnet and thus neglect the spin-dependence of $\dos$ and $\tau$.
 
The force and torque due to applied current are given as \cite{TKS_PR08}
\begin{eqnarray}
F & = & 
 -{\spol}\intx \, \nabla_{\xw}\evs \cdot\sev , \label{Fdef}
\\
\torqueve & = &
 -{\spol}\intx  (\evs \times\sev),\label{Tdef}
\end{eqnarray}
where $\sev$ is the electron spin density induced by the current and domain wall. 

The calculation of the elecron spin density is carried out by use of the spin gauge transformation, $a\equiv U c$, where 
$U$ is a $2\times2$ unitary matrix and $a$ is the electron operator in the gauge transformed frame. 
The marix $U$ is chosen to diagonalize the $s$-$d$ type interaction as $U\equiv \mv\cdot \sigmav$, where 
$\mv\equiv (\sin\frac{\theta}{2}\cos\phi,\sin\frac{\theta}{2}\sin\phi,
\cos\frac{\theta}{2})$ 
($\theta$ and $\phi$ are the polar coordinates of the localized spin direction, $\evs$).
This approach is justified if the adiabatic condition (\ref{adiabaticdisorder}) holds \cite{TKS_PR08}.

The spin-transfer torque and the force acting on a planar wall (with the wall plane perpendicular to the direction $\xw$) is expressed by the transverse spin densities, $\seth\equiv \sev\cdot\evth$ and $\seph\equiv \sev\cdot\evph$  
($\evth\equiv (\sin\theta\cos\phi,\sin\theta\sin\phi, \cos\theta)$ and $\evph\equiv (-\sin\phi,\cos\phi, 0)$)  as
\begin{eqnarray}
F =  - \spol \intx  (\nabla_\xw \theta) \seth 
   \nonumber\\
\torque_z  = -\spol \intx \sin\theta \seph.
\label{fandtorquedw}
\end{eqnarray}
Each component is expressed as
\begin{eqnarray}
\seth &=& 
 -\hf \sum_{\pm} e^{\mp i\phi} \stil^{\pm} \nonumber \\
\seph &=& 
-\hf \sum_{\pm} (\mp)ie^{\mp i\phi} \stil^{\pm},
    \label{sethph}
\end{eqnarray}
where 
\begin{equation}
\stil^{\pm}(\xv,t) \equiv  \average{\adag\sigma^\pm a},
\end{equation}
are the spin densities in the gauge-transformed frame, calculate by the standard diagramatical expansion.

\begin{figure}[tbh]
\begin{center}
\includegraphics[scale=0.3]{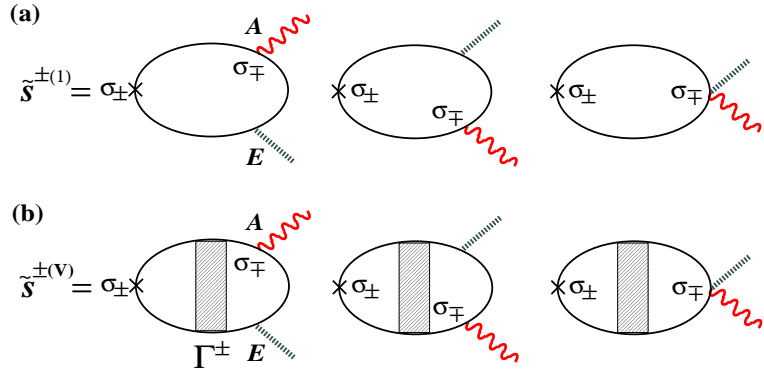}
\caption{ (a) Diagrammatic representation of the spin density
without vertex correction considered in Ref. \cite{TKS_PR08}.
(b) Vertex corrections to the spin density. 
Hatched square represents the diffusive ladder, $\Gamma^\pm$,  arising from successive electron scattering by the normal impurities.
\label{FIGspindensity}
}
\end{center}
\end{figure}

The spin density induced by the applied electric field $E$ was calculated in Ref.  \cite{TKS_PR08} without including the vertex correction. 
The result (denoted by $\stilpmo$) is
\begin{eqnarray}
\stilpmo_\qv  &=&
-\frac{e}{\pi ma^3}\sum_{ij} E_i A_j^\pm (\qv)
  I_{ij} ^\pm  (\qv) , \label{s1def}
\end{eqnarray}
where $a^3$ is the unit volume and 
\begin{eqnarray}
  I_{ij}^\pm  (\qv)\equiv 
\frac{1}{N} \sum_{\kv} 
\lt[
 \gr_{\kv-\frac{\qv}{2},\mp}  \ga_{\kv+\frac{\qv}{2},\pm}
\delta_{ij}
+\frac{\hbar^2 k_ik_j}{m}
 \gr_{\kv-\frac{\qv}{2},\mp} 
\lt( \gr_{\kv+\frac{\qv}{2},\pm} +\ga_{\kv-\frac{\qv}{2},\mp}  \rt)
 \ga_{\kv+\frac{\qv}{2},\pm}
\right],
\end{eqnarray}
where $N\equiv V/a^3$ is the number of sites and 
$\gr_{\kv,\mp}$ are the Green's function at zero frequency.
Here $A_\mu^\pm$ is the gauge field ($\mu$ and $\pm$ are the spatial and spin index, respectively). 
(Diagrams are shown in Fig. \ref{FIGspindensity} (a)).
In the adiabatic limit, $\kf \lambda \gg1$ and thus 
$I_{ij}^\pm  (\qv)$ can be approximated by the value at $q=0$, i.e.,  $I_{ij}^\pm  (0)$, since the transfer of the linear momentum between the gauge field and the electron can be neglected.

The aim of the present paper is to evaluate the vertex corrections, diagramatically shown in Fig. \ref{FIGspindensity} (b), which were not addressed to in Ref. \cite{TKS_PR08}.
%
%
%
%
The vertex correction contribution to the spin density is easily calculated as
\begin{eqnarray}
\stilpmv_\qv  &=&
-\frac{e}{\pi m a^3}\sum_{ij} E_i A_j^\pm (\qv)
  I_{ij}^{\pm} (\qv)  \Gamma^\pm(\qv) ,  \label{svdef}
\end{eqnarray}
where 
\begin{equation}
\Gamma^\pm(\qv) \equiv 
\sum_{n=1}^{\infty} (\nimp \vimp^2  I_{0}^\pm (\qv) )^n.
\label{Gammadef}
\end{equation}
Here 
\begin{equation}
  I_{0}^\pm (\qv)\equiv 
\frac{1}{N} \sum_{\kv}  \gr_{\kv-\frac{\qv}{2},\mp}  \ga_{\kv+\frac{\qv}{2},\pm}.
\end{equation}
is written also as
\begin{equation}
I_{0}^\pm = -\frac{1}{N}\sumkv (\gr_{\kv-\frac{\qv}{2},\mp} - \ga_{\kv+\frac{\qv}{2},\pm})\frac{1}{\pm2\spol +\frac{\hbar^2 \kv\cdot\qv}{m}+\frac{i\hbar}{\tau}} .
\label{Izero1}
\end{equation}

It was noted in Ref. \cite{TKS_PR08} that $\stilpmv$ is negligiblly small in ballistic 3$d$ ferromagnets, 
due to the strong spin splitting, $\spol\tau/\hbar\gg1$. 
This fact is easily checked by noting that 
$\nimp\vimp^2 I_{0}^\pm (\qv) =
\pm i \frac{\pi}{2}\nimp\vimp^2 \frac{\dos_++\dos_-}{\spol}+O(q^2)
=O\lt(\frac{\hbar}{\spol \tau}\rt) \ll1$ 
in this limit, and thus $\Gamma^\pm \ll1$.

In this paper, we are considering the opposite limit, 
$\spol\tau/\hbar\ll1$, namely, a dirty weak ferromagnet. 
We now demonstrate that the spin-transfer effect exists even in this case.
Mathematically, the dominant spin-transfer effect in this limit is included in the vertex correction, $\stilpmv$.
Expanding eq. (\ref{Izero1}) with respect to $q$ and $\spol\tau/\hbar\ll1$, we obtain
\begin{equation}
\nimp\vimp^2 I_{0}^\pm (\qv)=
\nimp\vimp^2 2\pi\dos\frac{\tau}{\hbar} \lt(1\pm2i\spol{\tau}/{\hbar} -Dq^2\tau\rt)
=\lt(1\pm2i\spol\tau/\hbar -Dq^2\tau\rt).
\label{Izero2}
\end{equation}
The summation in Eq. (\ref{Gammadef}) is then carried out as
\begin{equation}
\Gamma^\pm(\qv) = \frac{1}{Dq^2\tau \mp2i \spol \tau/\hbar}-1.
\label{Gammares}
\end{equation}
Therefore the total spin density including the vertex correction, $\stilpm\equiv \stilpmo +\stilpmv$,
 is obtained by use of Eqs. (\ref{s1def})(\ref{svdef})(\ref{Gammares}) as
\begin{eqnarray}
\stilpm_\qv  &=&
-\frac{e}{\pi m a^3}\sum_{ij} E_i A_j^\pm (\qv)
  I_{ij}^{\pm} (\qv) \frac{1}{Dq^2\tau \mp2i \spol \tau/\hbar} .  \label{svres1}
\end{eqnarray}
The last factor describes the long-range correlation of the torque induced by the diffusive electron motion. 
$I_{ij}^{\pm} (\qv)$ in the limit of 
$\spol\tau/\hbar\ll1$ is calculated as
\begin{align}
  I_{ij}^\pm  (\qv) &=
\frac{\delta_{ij}}{N} (\pm\spol)\sum_{\kv} 
\lt[  (\gr_{\kv})^2  \ga_{\kv} - \gr_{\kv}(\ga_{\kv} )^2  \rt]
\nnr
&=
\mp 4\pi i \dos\spol\frac{\tau^2}{\hbar^2} \delta_{ij} +o(q^2,\spol\tau/\hbar),
\label{Iijres}
\end{align}
where we have neglected the contribution containing higher order of $q$ and $\spol\tau$.
The expression for the gauge field in the case of a planar wall is given as \cite{TKS_PR08}
\begin{align}
A^\pm_\xw(\qv)= \mp i\frac{\pi}{2L}e^{\pm i\phi}\frac{1}{\cosh\frac{\pi}{2}q\lambda}\delta_{q_\perp,0},
\label{Apmres}
\end{align}
where $q$ represents the momentum transfer in the direction $\xw$ and $q_\perp$ represents that in the transverse direction and $L$ is the system length. 
From Eqs. (\ref{sethph})(\ref{svres1})(\ref{Iijres})(\ref{Apmres}), the  result of the spin polarization is obtained as
\begin{align}
\seph (q) &= 
\frac{3}{2}\pi 
\delta_{q_\perp,0} 
\frac{j}{e} \frac{\spol \tau}{ \ef \hbar L} 
\frac{1}{\cosh {\frac{\pi}{2}q\lambda}} 
\sum_{\pm} (\pm i) \frac{1}{Dq^2\tau\mp 2i\spol\tau/\hbar } 
\nonumber\\
&=
-6\pi 
\delta_{q_\perp,0} 
\frac{j}{e} \frac{(\spol \tau)^2}{\ef \hbar^2 L} 
\frac{1}{\cosh \frac{\pi}{2}q\lambda} 
\frac{1}{(Dq^2\tau)^2+4(\spol\tau/\hbar )^2} 
\\
\seth(q) &=
3\pi
\delta_{q_\perp,0} 
\frac{j}{e} \frac{\spol \tau}{\ef \hbar L} 
\frac{1}{\cosh\frac{\pi}{2}q\lambda} 
\frac{Dq^2\tau}{(Dq^2\tau)^2+4(\spol\tau/\hbar )^2} ,
    \label{sethphres}
\end{align}
where $j\equiv (e^2n\tau/m)E$ is the current density.
The final result of the torque and the force then becomes
(using $z\equiv \frac{\pi}{2}q\lambda$)
\begin{align}
\torque &= 
\frac{\pi^4 \hbar }{8} 
\frac{I}{e} \frac{\spol}{\ef} \lt(\frac{\lambda}{\lambdaD}\rt)^4 
\int_{-\infty}^\infty dz  
\frac{1}{\cosh ^2 z} 
\frac{1}{z^4 +\frac{\pi^4}{4}\lt(\frac{\lambda}{\lambdaD}\rt)^4 
} 
\\
F &=
\frac{3\pi^4 \hbar }{4} 
\frac{I}{e} \frac{\spol}{\ef \lambda } \lt(\frac{\lambda}{\lambdaD}\rt)^2
\int_{-\infty}^\infty dz  
\frac{1}{\cosh ^2 z} 
\frac{z^2}{z^4 +\frac{\pi^4}{4}\lt(\frac{\lambda}{\lambdaD}\rt)^4 
} .
    \label{torqueFres}
\end{align}

We are interested in the adiabatic limit, $\lambda \gg \lambdaD$, and the integrals are estimated in this limit as
\begin{align}
\int_{-\infty}^\infty dz  
\frac{1}{\cosh ^2 z} 
\frac{1}{z^4 +\frac{\pi^4}{4}\lt(\frac{\lambda}{\lambdaD}\rt)^4 
}
 & \sim 
\frac{4}{\pi^4} \lt(\frac{\lambdaD}{\lambda}\rt)^4 
\int_{-\infty}^\infty dz  
\frac{1}{\cosh ^2 z}  
= \frac{8}{\pi^4}\lt(\frac{\lambdaD}{\lambda}\rt)^4 
\\
\int_{-\infty}^\infty dz  
\frac{1}{\cosh ^2 z} 
\frac{z^2}{z^4 +\frac{\pi^4}{4}\lt(\frac{\lambda}{\lambdaD}\rt)^4 
}
 & \sim 
\frac{4}{\pi^4} \lt(\frac{\lambdaD}{\lambda}\rt)^4 
\int_{-\infty}^\infty dz  
\frac{z^2}{\cosh ^2 z}  
= \frac{2}{3\pi^2}\lt(\frac{\lambdaD}{\lambda}\rt)^4 
\end{align}

The torque and the force in the disordered adiabatic limit is 
finally obtained as
\begin{align}
\torque &= 
\frac{\hbar I}{e} \frac{\spol}{\ef} 
\\
F &=
\frac{\pi^2}{2} 
\frac{\hbar I}{e} \frac{\spol}{\ef \lambda } 
\lt(\frac{\lambdaD}{\lambda}\rt)^2
    \label{torqueFfinalres}
\end{align}
What is significant is that the result of the torque indicates that the transfer of the spin angular momentum is essentially 100\% in the disordered adiabatic limit. 
In fact, if we define the spin polarization in the disordered case as
\begin{align}
P_{\rm D}\equiv \frac{\spol}{\ef},
\end{align}
we see that the torque is simply given as
\begin{align}
\torque=\frac{\hbar I}{e}P_{\rm D}.
\label{torqueP}
\end{align}
In the ballistic adiabatic limit, on the other hand, we know that the spin-transfer torque is given by 
$\torque=\frac{\hbar I}{e}P$, where the polarization is defined as
$P\equiv \frac{n_+-n_-}{n_++n_-}$ ($n_\pm$ is the density of the electron with spin $\pm$) \cite{TKS_PR08}.
Since $P_{\rm D}\sim P$ in most cases, we see that Eq. (\ref{torqueP}) indicates that the perfect spin transfer occurs even in the disordered weak ferromagnets as long as the disordered adiabatic condition 
$\lambda \gg \lambdaD$ is satisified. 
One should note that the actual magnitude of the spin transfered in a weak ferromagnet is small, proportional to the small polarization factor, $P_{\rm D}$.
Nevertheless, the efficienty of the wall motion is as high as in strongly polarized ferromagnets, as we will demonstrate below.

The force on domain wall can be measured by a dimensionless parameter $\beta$ defined as
$\beta \equiv \frac{eS\lambda}{\hbar I} F$ \cite{TKS_PR08}, where $S$ is the magnitude of localized spin.
From Eq. (\ref{torqueFfinalres}), the parameter $\beta$ arising from the difussive electron motion is given by 
\begin{align}
\beta = \frac{\pi^2}{2}S P_{\rm D} 
\lt(\frac{\lambdaD}{\lambda}\rt)^2.
    \label{beta}
\end{align}
We see here that $\lt(\frac{\lambdaD}{\lambda}\rt)^2$ is a measure of the non-adiabaticity.
If $P_{\rm D}\sim0.1$ and $\lambdaD/\lambda\sim0.2$ with $S\sim1$, we obtain $\beta=0.02$, which 
is sufficiently large to improve the wall motion greatly \cite{Thiaville05,TKS_PR08}.

\begin{figure}[tbh]
\begin{center}
\includegraphics[scale=0.5,angle=270]{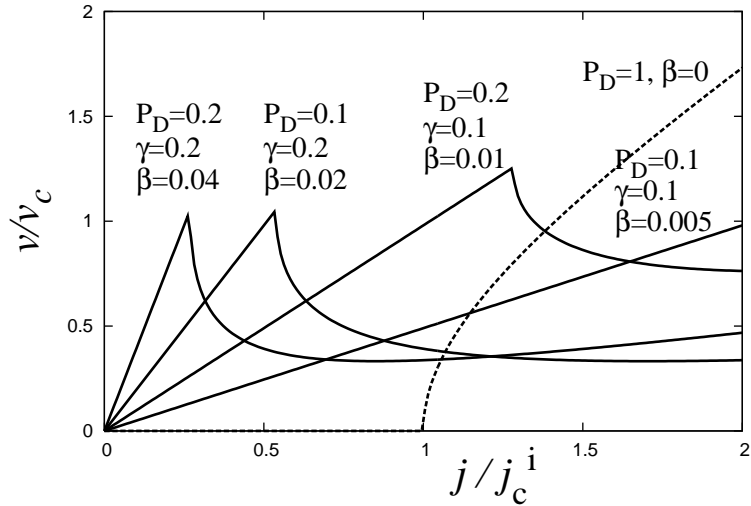}
\caption{ 
The wall velocity as function of the applied current density for different values of $P_{\rm D}$ and $\gamma\equiv \lt(\frac{\lambdaD}{\lambda}\rt)$ (solid lines).
The case of the ballistic adiabatic limit with $\beta=0$ is plotted by a dotted line for comparison.
The velocity and the current density are normalized by $\vc$ and the intrinsic threshold current 
($j_{\rm c}^{\rm i}=\frac{2eS\vc}{a^3}$) 
at $P_{\rm D}=1$, respectively.
\label{FIGdwvj_diffusive}
}
\end{center}
\end{figure}

Let us study the wall dynamics in the present diffusive regime.
We neglect the extrinsic pinning.
The equation of motion of a planar wall is given by \cite{TKS_PR08}
\begin{eqnarray}
\dot\phiz+\alpha \frac{\dot{X}}{\lambda}
 &=& \frac{a^3}{2eS\lambda} \beta j \nonumber\\
\dot{X}-\alpha\lambda\dot{\phiz} &=&  \vc \sin 2\phiz  
+\frac{a^3}{2eS}P_{\rm D}{j},
\label{DWeq3}
\end{eqnarray}
where 
$\vc\equiv \frac{\Kp\lambda S}{2\hbar}$ is a critical velocity corresponding to the hard axis anisotropy energy $\Kp$. 
From Eq. (\ref{DWeq3}), we immediately notice that the intrinsic threshold, given as $\frac{2eS\vc}{P_{\rm D}a^3}$ is very high for a small spin polarization. 
Nevertheless, the wall motion occurs due to the $\beta$ term induced by the diffusive motion.
The velocity calculated as function of the applied current density for different values of $P_{\rm D}$ and 
$\gamma\equiv \lt(\frac{\lambdaD}{\lambda}\rt)$ is plotted by solid lines in Fig. \ref{FIGdwvj_diffusive}.
It is seen that the wall motion indeed occurs at lower current density even compared with the the intrinsic threshold at 100\% spin polarization ($j_{\rm c}^{\rm i}=\frac{2eS\vc}{a^3}$).
We also see that the wall motion is governed by the parameter $\beta$ and not much by the polarization $P_{\rm D}$ in the present diffusive regime.
Therefore, strogly polarized ferromagnet is not the necessary condition for efficient domain wall motion, but disordered weak ferromagnet is another option as promising as strong ferromagnets.

To conclude, we have derived the adiabatic condition for the spin transfer effect in disordered weak ferromagnets (Eq. (\ref{adiabaticdisorder})), and showed that the perfect spin transfer effect occurs if that condition is fullfilled even in the case of weak $s$-$d$ type coupling. 
We have also derived the force acting on the wall in the diffusive limit and estimated the corresponding $\beta$, which turned out to govern the wall motion in the diffusive weak ferromagnets.
By solving the equation of motion of a planar wall under current, we found that wall motion as efficient as strong ferromagnets can be realized in the present system, due to a large value of force or $\beta$. 

Our result also serves as a proof that the gauge field expansion is justified as long as disordered adiabatic condition (Eq. (\ref{adiabaticdisorder})) is satisfied.

This work was supported by a Grant-in-Aid for Scientific Research in Priority Areas, "Creation and control of spin current" (1948027), the Kurata Memorial Hitachi Science and Technology Foundation and the Sumitomo Foundation.

%
\end{document}